\documentclass{nature_meth}
\usepackage{amssymb,amsfonts,amsmath}
\usepackage{graphicx} 
\usepackage{gensymb}

\usepackage{epsfig}
\usepackage[figoff]{figcaps}

\title {Which Contrast Does Matter?   Towards a Deep Understanding of MR Contrast using Collaborative GAN}
\author{Dongwook Lee$^1$, Won-Jin Moon$^2$ \& Jong Chul Ye$^{*,1}$
}

\begin{document}
\maketitle

\begin{affiliations}
\item Department of Bio and Brain Engineering, KAIST, Daejeon, Korea
\item Konkuk University Medical Center, Seoul, Korea
\item[] Correspondence should be addressed to J.C.Y. (jong.ye@kaist.ac.kr)
\end{affiliations}

%\setstretch{2}
\begin{abstract}
\baselineskip 0.23in

Thanks to the recent success of generative adversarial network (GAN) for image synthesis,
there are many exciting GAN approaches that successfully synthesize MR image contrast from other  images with different contrasts.
These approaches are potentially important for  image imputation problems, where  complete set of data is often difficult to obtain and image
synthesis is one of the key solutions for handling the missing data problem.
Unfortunately,  the lack of the scalability of the existing GAN-based image translation approaches poses a fundamental challenge to understand the nature of the MR contrast imputation problem:  which contrast does matter? 
%In the age of artificial intelligence (AI), clinicians are interested in understanding which MR contrast is really in- dispensable and cannot be synthesized using a generative model. 
%This is especially important for clinical decision making, since there are many claims that they can successfully synthesize any MR contrast. It is generally believed that the success of MR contrast is due to the redundancies in the different MR contrast. Therefore, 
%To understand the nature of MR contrast imputation, we should understand the redundancies across many different MR contrasts to figure out what kinds of information can be, or cannot be generated. 
%However, such an analysis is not trivial, since the understanding of the redundancies across multiple MR contrasts requires complete knowledge of joint image manifolds, which is considered as complicated machine learning task.
%Many problems in medical imaging require a set of multiple images, but the complete set of data is often difficult to obtain. Image imputation is one of the key solutions for handling the missing data problem.
%Especially in MR contrast imputation problems, the collaborative processing of the multiple input images is very important to find the target contrast image with accurate pixel intensity. 
Here,  we present a systematic approach using  Collaborative Generative Adversarial Networks (CollaGAN), which enable
the learning of  the joint image manifold of multiple MR contrasts to investigate which contrasts are essential.
Our experimental results showed that the exogenous contrast from contrast agents is not replaceable,
but other endogenous contrast such as T1, T2, etc can be synthesized from other contrast.
These findings may give important  guidance to the acquisition protocol design for MR  in real clinical
environment.

%We demonstrated that CollaGAN reconstructs the specific contrast MR images without any artifacts. CollaGAN can reduce the scan time by avoiding the additional scan for accurate clinical diagnosis.
%According to our brain tumor segmentation study, we have also shown that the images with the contrast agent injection cannot be imputed by generative models.
%Furthermore, CollaGAN can be used for all other types of imaging studies to investigate which contrast is essential and which contrast is redundant.

\end{abstract}
 
\newpage
%\setstretch{2}

In many image processing and computer vision  problems in medical imaging,
a set of multiple images are usually required to find a desired output.
%For example, in order to render a 3-D volume from  multiple view images, most algorithms require a set of images from the pre-defined view angles~\cite{choy20163d}.
%In medical imaging applications, 
For example,  for accurate diagnosis and segmentation of the cancer margin and  radiomic evaluations,
%magnetic resonance images (MR images) with various contrasts such as T1-weighted image, T1-weighted image with contrast agent, T2-weighted image and T2-FLAIR (FLuid-Attenuated Inversion Recovery) are usually acquired
% ~\cite{drevelegas2011imaging}. 
%More specifically,  for brain cancer diagnosis and radiomics studies,
multiple MR constrast such as T1-weighted (T1), post-contrast T1-weighted (T1Gd), T2 weighted (T2), and T2-FLAIR (T2F) are necessary \cite{drevelegas2011imaging,menze2015multimodal,bakas2017advancing}. 
Unfortunately, the complete set of input data are often difficult to obtain due to the different acquisition protocol at each institute, 
prolonged acquisition time, operator errors, or patient movement during the data acquisitions. Moreover,  it it often impossible to use  contrast agents for some patients with  kidney failure or allergic responses.
Without the complete contrast, the subsequent analysis can be prone to substantial biases and  errors that  can reduce the statistical efficiency of subsequent analysis~\cite{baraldi2010introduction},
%
% When some of the contrast
%are missing, 
and the accurate segmentation of the whole tumor, tumor core and effective tumor core may not be feasible.

Moreover, in some situations, although  multiple contrast images are available, some of the images suffers from systematic errors.
For example, 
a synthetic MRI technique called Magnetic Resonance Image Compilation (MAGiC, GE Healthcare)~\cite{tanenbaum2017synthetic}  enables the generation of the various contrasts MR images 
using a
Multi-Dynamic Multi-Echo (MDME) scan. While MAGiC can provide clinically useful synthetic MR images with various contrasts such as T1-weighted, T2-weighted, T2-FLAIR, etc, it is often reported that some of the synthetic contrasts have readily recognizable artifacts ~\cite{tanenbaum2017synthetic, hagiwara2017synthetic, hagiwara2017symri}. 
Especially, the characteristic granulated hyperintense artifacts apparent in the margins along the cerebrospinal fluid (CSF)-tissue boundaries on MAGiC FLAIR can be mistaken for true pathologic conditions such as meningeal disease or subarachnoid hemorrhage in clinical practice. Furthermore, flow and/or noise artifacts are more frequent on MAGiC FLAIR than conventional FLAIR. 
This often leads to the additional MR acquisition to confirm the diagnosis, 
%and
%the patient are often undergone additional MR scans, 
which requires significant amount of cost and patient inconvenience.

Therefore,  rather than re-acquiring all data as a complete set in this unexpected situation, 
%which is generally not feasible,
 it is often necessary to fill the missing data with substituted data. In statistical literatures,
this process is often referred to as \textit{missing data imputation}. Once all missing data have been imputed, the dataset can be used as an input for standard techniques designed for the complete dataset.

Recently, the field of image imputation has been significantly advanced thanks to the 
enormous success of deep neural networks \cite{krizhevsky2012imagenet, zhang2017beyond, dong2016image, xie2012image,deng2009imagenet}.
Typically, the missing image imputation
problem can be formulated as an  image translation problem  from one domain to the other domain \cite{zhu2017unpaired, choi2017stargan} ,
whose performance has been  great improved with the advance of
 Generative Adversarial Network (GAN)~\cite{goodfellow2014generative}.
 %  these approaches of computer vision applications are significantly improved~\cite{ledig2017photo, yeh2017semantic}.
The main purpose of GAN architecture is to generate the realistic samples/images. 
Typical GAN consists of two neural networks: a generator and a discriminator.
The discriminator attempts to find the features to distinguish fake image from real images, while the generator learns to synthesize images so that the discriminator is difficult to judge as real or fake.
After training both neural networks, the generator produces the realistic outputs which cannot be distinguished as fake samples by the discriminator.
Since the introduction of the original GAN ~\cite{goodfellow2014generative}, many ingenious extensions have appeared.
For example, 
for the translation between two domains $A$ and $B$,  CycleGAN \cite{zhu2017unpaired} constructs two generators, $G_{A\rightarrow B}$ and $G_{B\rightarrow A}$, and two discriminators, $D_A$  and $D_B$, so that
the images between two domains can be successfully translated  by cycle consistency loss~\cite{wolterink2017deep}.
In another variation, to handle the multiple domains more than two,
 Choi et al. proposed StarGAN~\cite{choi2017stargan} which utilized the shared feature learning using a single generator and a single discriminator. Using the concatenated input image with target domain vector, the generator produces the fake image, which is classified as the target domain by the discriminator.

Inspired by the success of GAN-based image-translation techniques, there have been many attempt to generate
MR contrast. For example, Dar et al. proposed MR contrast synthesis with conditional GAN and additional perceptual loss~\cite{dar2019image}. Specifically, they utilized Pix2pix~\cite{isola2017image} and CycleGAN to translate the MR contrast images between T1 and T2 weighted images. Welander et al. compare the performance of CycleGAN and UNIT~\cite{liu2017unsupervised} in the task of translation between T1 and T2 weighted images~\cite{welander2018generative}.
Furthermore, there are severaly studies to translate the images between MR and CT (Computed Tomography) by using
similar cycle consistency loss~\cite{yang2018unpaired,wolterink2017deep,hiasa2018cross}. 
%They utilized the unpaired image-to-image translation concept from CycleGAN to translate the images from the other medical imaging modalities.
Meanwhile, Hagiwara et al. proposed the conditional GAN-based frameworks to generate desired FLAIR images by two step approach~~\cite{hagiwara2019improving}. First, FLAIR images are generated by MAGiC. Since the MAGiC FLAIR have synthetic artifacts, they tried to remove the synthetic artifact and  improve the quality of synthetic MR imaging by utilizing Pix2pix after the MAGiC. %However, it has limited application since it focuses only on the synthetic FLAIR imaging.

Despite of this success, handling the multiple inputs is one of the challenges for existing image-to-image translation approaches. For example, 
 for translating among $N$ number of domains using CycleGAN, it is necessary to train $N(N$-$1)$ generators for each
 pair of domains,  and $N$ discriminators for each domain. Therefore, CycleGAN requires a large number of neural networks in multi-domain setting since it is trained without feature sharing among the multi-domains.
 Although StarGAN ~\cite{choi2017stargan} address the multiple domain mapping,  %Similarly, to generate a missing contrast,  
 it 
 cannot exploit the redundancies across MR contrast images to reconstruct the output contrast,
  since StarGAN is designed to utilize  only one input.
 
In fact, this lack of the scalability of the existing GAN-based image translation approaches poses a fundamental challenge
to understand the nature of the MR contrast imputation problem: i.e. which contrast does matter?
In the age of artificial intelligence (AI) with amazing success of generative models, 
clinicians are interested in understanding which MR contrast is really indispensable and cannot be synthesized using 
a generative model. This is especially important for clinical decision making, since there are many claims that they can successfully
synthesize any MR contrast.
It is generally believed that the success of MR contrast is due to the redundancies in the different MR contrast. 
Therefore, to understand the nature of MR contrast imputation, we should understand the redundancies across many
different MR contrasts  to figure out what kinds of information can be, or cannot be generated.
However, such an analysis is not trivial, since the understanding of the redundancies
across multiple MR contrasts requires complete knowledge of
 joint image manifolds, which  is considered as complicated machine learning task.
% and it is not clear how to investigate the
%combined image manifold across multiple images.

To address general image imputation problems in computer vision and image processing,  we recently developed a novel  image imputation method called Collaborative Generative Adversarial Network (CollaGAN) \cite{lee2019collagan},  which reconstruct the missing image by learning the redundancies across many  image pairs.
In CollaGAN,
%addresses the aforementioned challenges by treating the images-to-image translation task as the missing image imputation problem. 
a set of images from the whole domains is treated as a complete set, 
% and the desired domain image to reconstruct is treated as a missing image to impute.
and  the network is trained to estimate  the missing image by synergistically combining the information from the multiple inputs.
The power of the method has been successfully demonstrated to generate facial expressions, lightning conditions, etc \cite{lee2019collagan}.

\begin{figure}[!h]     
\centerline{\epsfig{figure=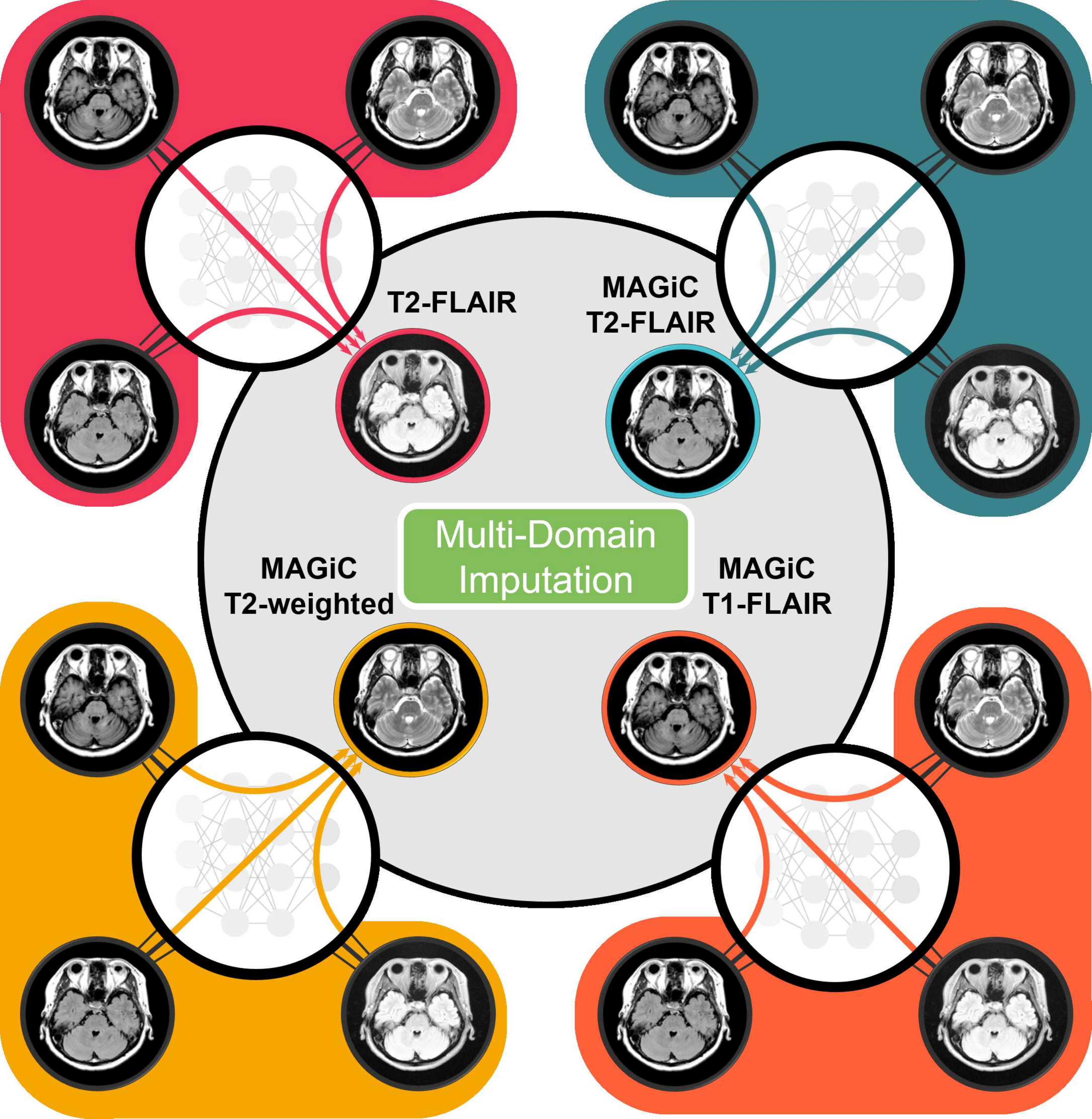, width=0.8\linewidth}}
 \centering
   \caption{\bf\footnotesize Concept diagram of multi-domain imputation task on MR contrast images using the proposed Collaborative GAN (CollaGAN). The single generator (black circle) of CollaGAN utilizes multiple input images from various contrasts to synthesize the target contrast image. Here, T1-FLAIR (orange circle), T2-weighted (yellow circle), MAGiC T2-FLAIR (green circle) from synthetic MR imaging (MAGiC) and the additional scan by conventional T2-FLAIR (red circle) were used for multi-domain imputation. }
 \label{fig:concept_colla}
\end{figure}

Inspired by this success, Fig~\ref{fig:concept_colla} illustrates how CollaGAN can be used for the case of MR contrast imputation problems.
In particular,  the collaborative processing of the multiple domain input images is very important in MR contrast imputation problems, since it is impossible to find the accurate pixel-intensity without understanding image manifold
across different contrast. 
This may appear similar to MAGiC that  calculates the voxel intensity from multi-contrast MR images from a MDME scan.
However, in contrast
to MAGiC, the collaborative learning with CollaGAN also utilizes the semantic information beyond the pixel wise relationship, so the more systematic studies about the MR contrast can be performed.
%, and we could obtain
%%Utilization of the multiple contrast images as inputs collaboratively, is a necessary step for the 
%accurate reconstruction of the desired MR contrast  as shown in Fig.~\ref{fig:res_comparison}.
%
%MAGiC also calculates the voxel intensity from the multi-contrast MR images from the MDME scan.
Moreover, unlike CycleGAN, CollaGAN utilizes a single discriminator and a single generator to reconstruct the image of whole domains so that
 the generator can effectively exploit the multiple domain redundancy by learning
 high dimensional manifold structure across images.
% shares the features from each domain while the training. It helps to have scalability for CollaGAN compared to CycleGAN.
%Especially, the collaboration of the multiple domain is very important in MR contrast imputation problem since it is impossible to find the accurate pixel-intensity from single contrast image. 
Specifically, by estimating specific contrast from the rest, we can understand the joint manifold structure across multiple contrast to decide
which contrast is most essential and  cannot be generated effectively. 
This is very important in clinical environment, since one can reduce the unnecessary exams while retaining the most essential one.

To validate the use of CollaGAN in understanding the essential MR contrast, we first perform quantitative study by comparing the segmentation performance by replacing one real contrast with a synthesized contrast. 
Here, we utilized the multi-modal brain tumor image segmentation benchmark (BraTS, 2015)~\cite{menze2015multimodal,bakas2017advancing}. 
All the scans from BraTS consist of T1-weighted (T1), post-contrast T1-weighted (T1Gd), T2 weighted (T2), T2-FLAIR (T2F) and the ground truth segmentation labels for brain tumors.
The segmentation performances were evaluated from the five different BraTS datasets: $Original$, $T1_{Colla}$, $T1Gd_{Colla}$, $T2_{Colla}$, and $T2F_{Colla}$. The datasets with subscript $_{Colla}$ represent the datasets with the substitution of a specific contrast by the reconstructed contrast from CollaGAN.
Here, for brain tumor segmentation, we used the state-of-the-art segmentation network known as convolutional neural network  with variational auto-encoder regularization~\cite{myronenko20183d} with some minor modifications. 

\begin{figure}
\centering
\centerline{\epsfig{figure=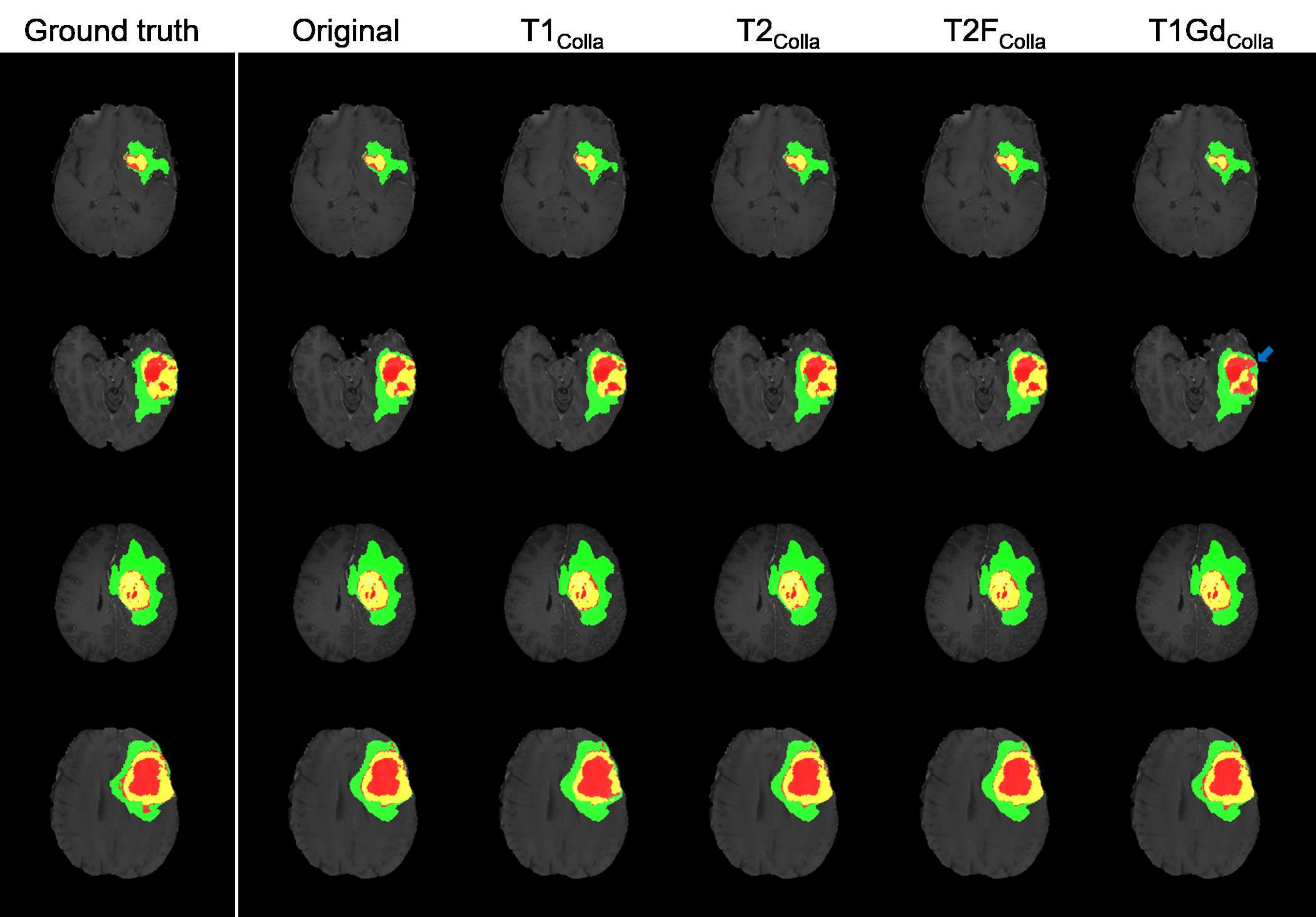, width=1\linewidth}}
\caption{\bf\footnotesize BRaTS segmentation results for quantitative evaluation of CollaGAN. 
The labels of the whole tumor (union of green, red and yellow), the tumor core (union of red and yellow), and the enhancing tumor core (yellow) are overlaid on T1-weighted sagittal images.
The segmentation results from the original BRaTS datasets (2nd row) show similar to the ground truth (1st row).
 $T1_{Colla}$, $T2_{Colla}$, $T2F_{Colla}$, and $T1Gd_{Colla}$ represent the datasets which the T1 weighted images, T2 weighted images, T2-FLAIR, and T1 weighted images with Gd-injection are respectively substituted by the reconstructed images from CollaGAN. For whole tumor, tumor core and enhancing tumor core, it shows the similar segmentation performance on both the original BRaTS and the reconstructed BRaTS datasets from CollaGAN. In $T1Gd_{Colla}$ (last row), the prediction of enhancing tumor core shows inferior performance (blue arrows) since $T1Gd_{Colla}$ has the lack of the information from the Gd-injection which is necessary to predict accurate enhancing tumor core region. } 
\label{fig:res_seg2}
\end{figure}

Fig.~\ref{fig:res_seg2} shows the segmentation results for the five different BraTS datasets. As shown in Fig.~\ref{fig:res_seg2}, the segmentation network performs well to find the whole tumor (WT), the tumor core (TC), and the enhancing tumor core (EC) maps on the original BraTS dataset. The segmentation maps from the synthetic BraTS datasets ($T1_{Colla}$, $T2_{Colla}$, $T2F_{Colla}$ and $T1Gd_{Colla}$) produced similar results to the ground truth and the result maps from the original BraTS.
For quantitative evaluation, we measured the segmentation performance by the dice similarity score~\cite{dice1945measures} between the prediction map, $Y_{pred}$, and the ground truth, $Y_{gt}$:
$$ Dice(Y_{gt},Y_{pred}) = \frac{2|Y_{gt}\cap Y_{pred}|}{|Y_{gt}|+|Y_{pred}|},$$
where $|\cdot|$ represents the cardinalities of the set (number of elements in each set).
The segmentation network achieves 0.8531$\pm$0.0869 / 0.7368$\pm$0.1850 / 0.7066$\pm$0.2717 (mean$\pm$std, $N$=28) DICE scores for WT / TC / EC, respectively, when using the original BraTS datasets.
When the original T1-weighted images are replaced with the reconstructed image by CollaGAN ($T1_{Colla}$), the DICE scores are reaches to 0.8567$\pm$0.0882 / 0.7342$\pm$0.1857 / 0.6979$\pm$0.2718 for WT / TC / EC, respectively, without any additional training or fine-tuning process.
The segmentation performance between the original and ($T1_{Colla}$,  $T2_{Colla}$, and $T2F_{Colla}$) are very similar as shown in Fig.~5. The results validated the reconstructed contrast images by CollaGAN for the data set $T1_{Colla}$,  $T2_{Colla}$, and $T2F_{Colla}$ are very similar to the original contrast images from the original BraTS dataset.
%The segmentation performances are maintained for every $T1_{Colla}$, $T1Gd_{Colla}$, $T2_{Colla}$, and $T2F_{Colla}$ dataset as shown in Fig~\ref{fig:res_seg}.

\begin{figure}%[h!]%  figure placement: here, top, bottom, or page
   \centering
\centerline{\epsfig{figure=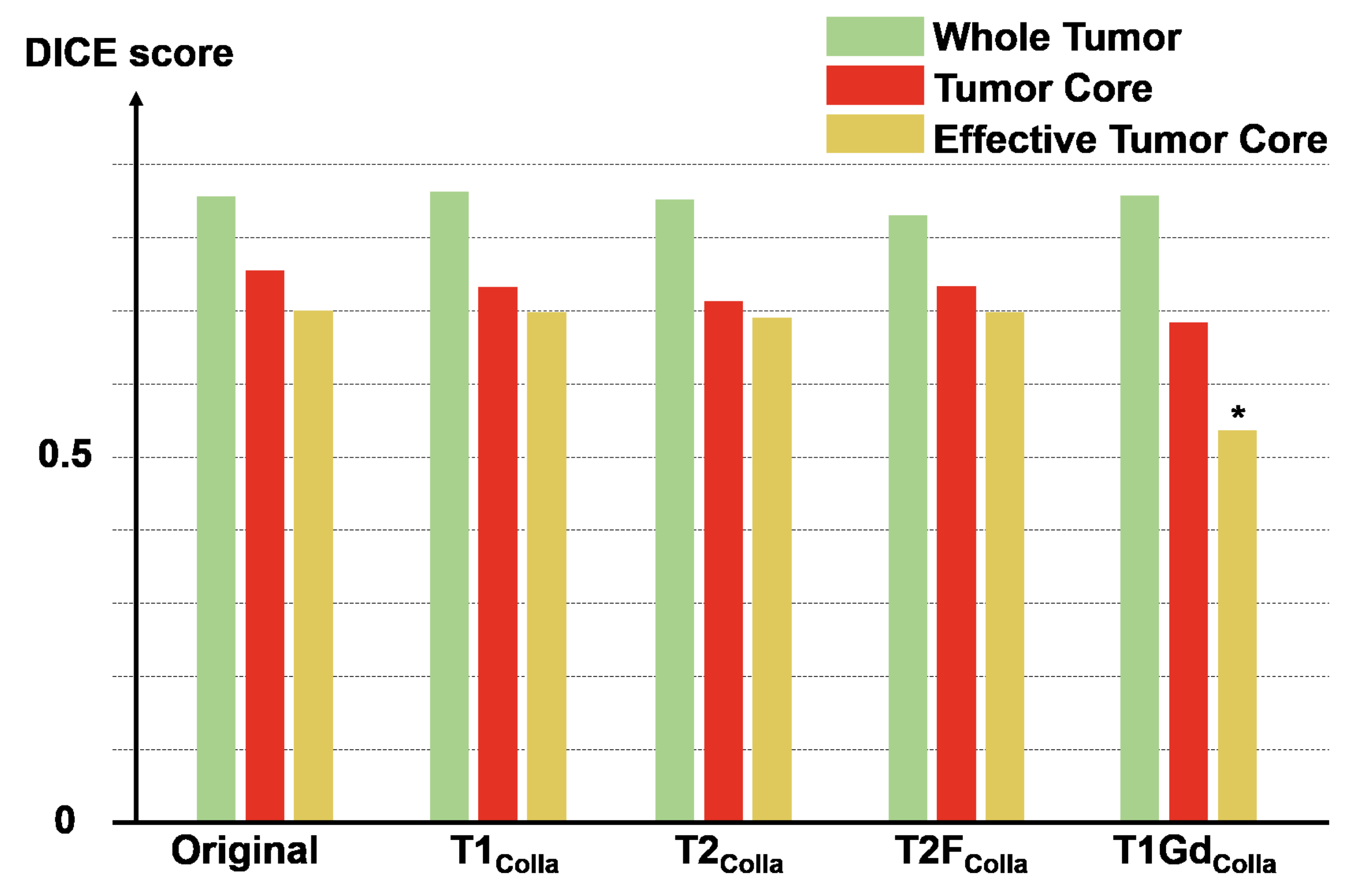, width=0.7\linewidth}}
 \vspace*{0.5cm}
   \caption{\bf\footnotesize Quantitative results for segmentation performance using the following datasets: original BRaTS and 
 $T1_{Colla}$, $T2_{Colla}$, $T2F_{Colla}$, and $T1Gd_{Colla}$ data set. Here,  $T1_{Colla}$, $T2_{Colla}$, $T2F_{Colla}$, and $T1Gd_{Colla}$  represent the datasets which the T1 weighted images, T2 weighted images, T2-FLAIR and T1 weighted images with Gd-injection are respectively substituded by the reconstructed images from CollaGAN. The segmentation network shows similar performance for the whole tumor, the tumor core, and enhancing tumor core on both the original BRaTS and the reconstructed datasets by CollaGAN for  $T1_{Colla}$, $T2_{Colla}$, and $T2F_{Colla}$ data set.
 However, the prediction of enhancing tumor core shows inferior performance in $T1Gd_{Colla}$,  since $T1Gd_{Colla}$  is lack of the information from the Gd-injection which is necessary for accurate prediction of the enhancing tumor core region. }
 \label{fig:res_seg}
\end{figure}

However, the injection of the gadolinium contrast agent provides additional tissue information so that the post-contrast T1-weighted (T1Gd) images show important role in the segmentation of  TC and EC as shown in the performance
drop of the segmentation results using $T1Gd_{Colla}$.
While the performance drop from  CollaGAN reconstructed T1Gd images using the other contrasts is relatively
small  for WT and TC  in Fig.~\ref{fig:res_seg},  the performance drop in EC is statistically significant.
%since the accurate reconstruction of T1Gd images  only using the other contrast images with no information from contrast agent is not feasible without additional information. %le contrast translation task and it is very difficult.
%is not just a contrast translation of MR tissues, but a provider for additional information to segment EC and TC given by the injection of Gadolinium contrast agent. 
%In synthetic MRI, therefore, it is impossible to synthesize T1Gd images due to the lack of tissue contrast information given by Gadolinium injection.
This experimental provides systematic understanding that the information of contrast injection is still indispensable unless
additional diagnostic evaluation is performed. This is as expected given the wide use of MR contrast agent.

\begin{figure}[bt]
\centering
\centerline{\epsfig{figure=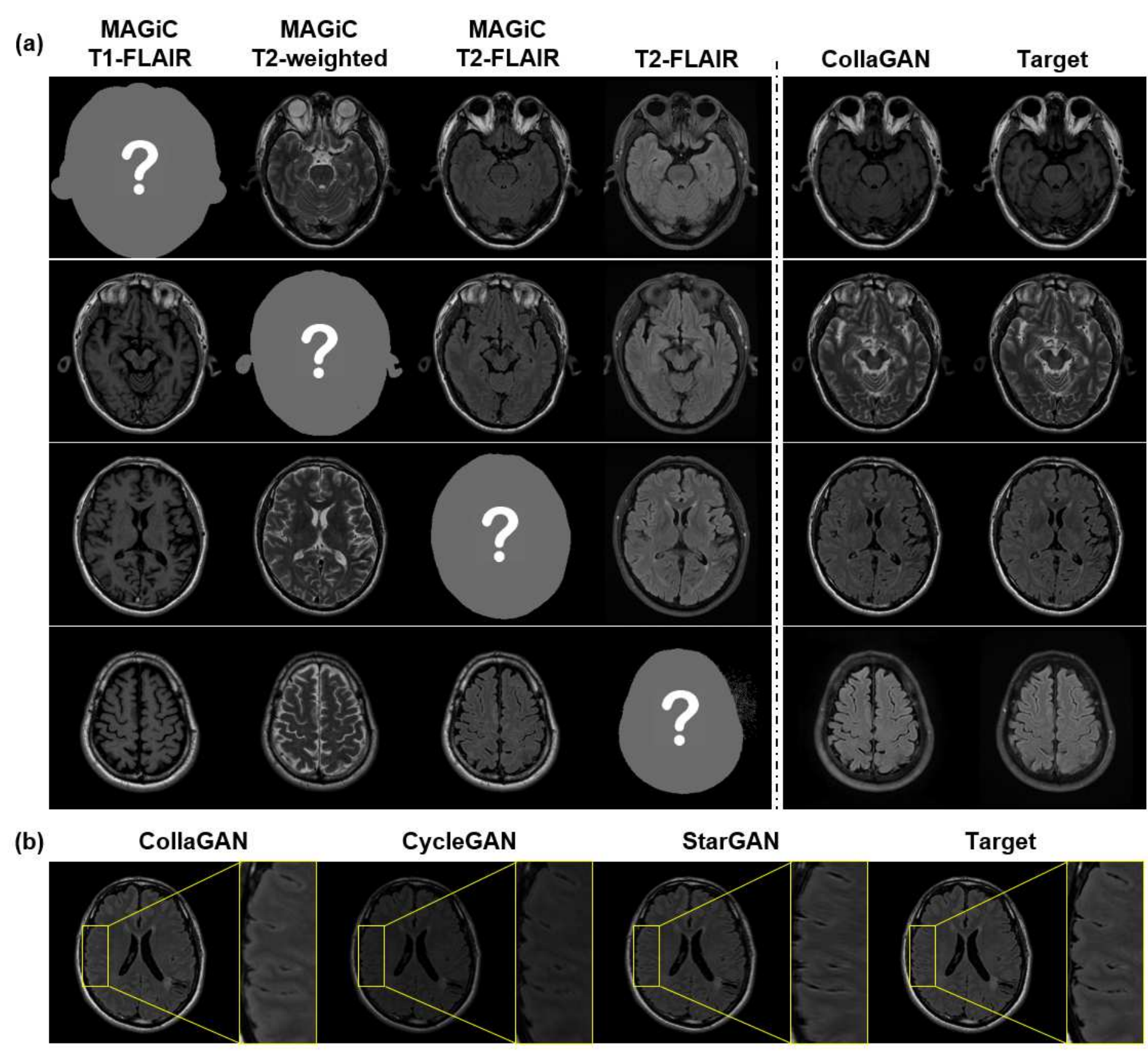, width=1\linewidth}}
\caption{\bf\footnotesize (a) MR contrast imputation results  using the proposed method, CollaGAN. The target domain contrast images (right) were reconstructed from the other contrast inputs (left).  
The contrast image to impute is marked as the question mark.
The normalized mean squared error (NMSE) and the structural similarity index (SSIM) of the results are as following (NMSE/SSIM): 0.0326/0.918 for T1-FLAIR, 0.109/0.904 for T2-weighted, 0.0238/0.942 for MAGiC T2-FLAIR, and 0.110/0.740 for T2-FLAIR. (b) Comparison of T2-FLAIR imputation results using CycleGAN, StarGAN and CollaGAN with $\times$3 magnified images.
 }
\label{fig:res_comparison}
\end{figure}

\begin{figure}
\centering
\centerline{\epsfig{figure=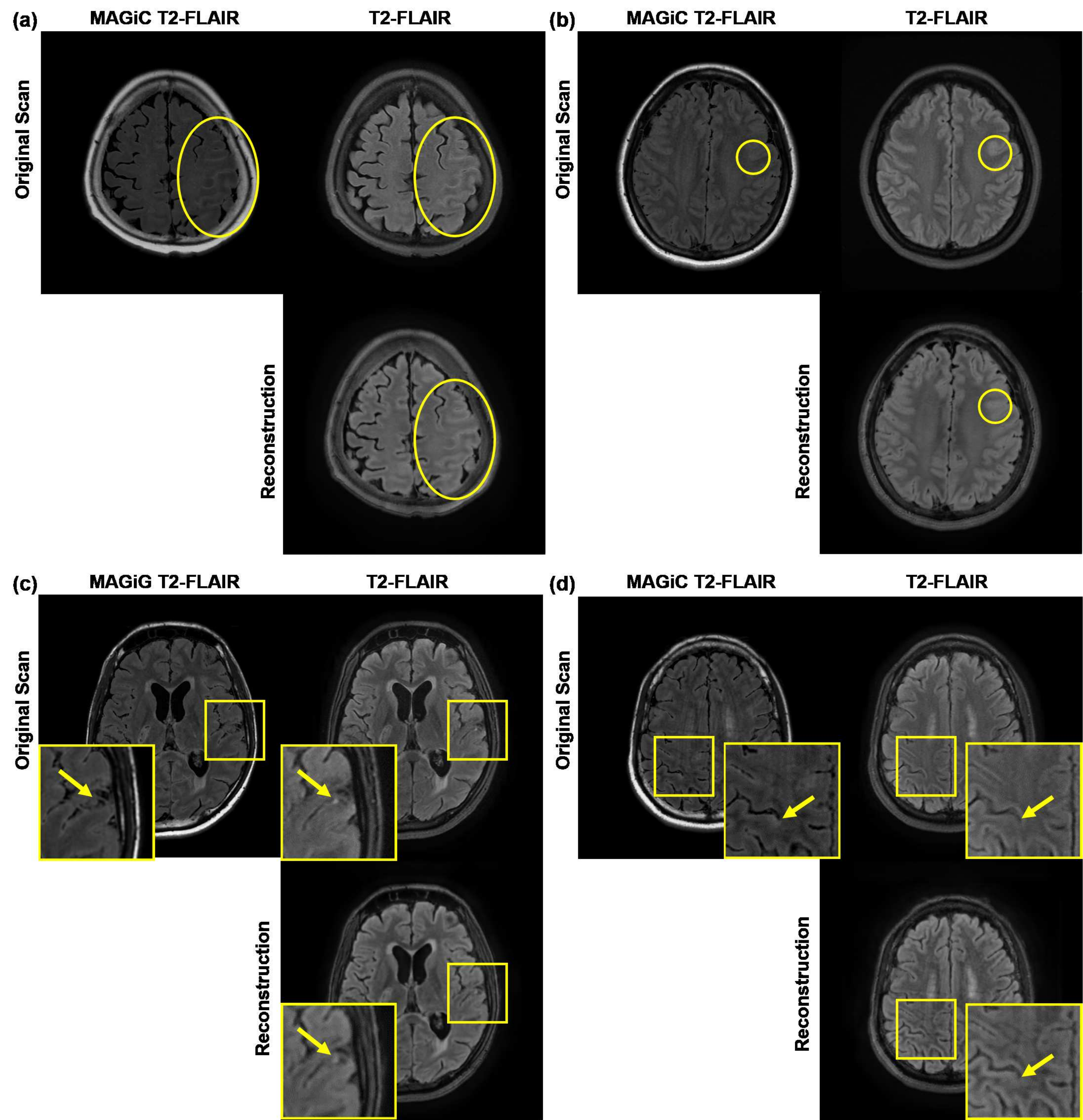, width=0.7\linewidth}}
\caption{\bf\footnotesize Reconstruction results with lesions by CollaGAN.  
MAGiC T2-FLAIR,  the ground truth (T2-FLAIR, separated conventional scan), and the reconstructed T2-FLAIR using CollaGAN are illustrated.
When there is no synthetic artifact on MAGiC T2-FLAIR (a-b), the reconstructed contrasts from CollaGAN also show the accurate performance. (a) The hyperintensity signal of the CSF space (circled) compared to the other side hemisphere is visible on all three images. (b) The cortical and sulcal abnormality (circled) is also visible on all three images. 
On the other hand, although there exists synthetic artifact on MAGiC T2-FLAIR (c-d), CollaGAN  reconstructs the artifact-free T2-FLAIR results similar to the ground-truth T2-FLAIR images.
(c) The focal sulcal hyperintensity (arrow) is only visible on original T2-FLAIR and reconstruction T2-FLAIR by CollaGAN,
while it is not visible on the MAGiC T2-FLAIR images. The MAGiC T2-FLAIR cannot capture the hyperintensity lesion, but CollaGAN can accurately capture it.
(d) In MAGiC T2-FLAIR, there is a pseudo-lesion (arrow) which is not visible on both original T2-FLAIR and reconstructed T2-FLAIR by CollaGAN.
}
\label{fig:res_lesion}
\end{figure}

Although the previous experiment shows that the gadolinium contrast cannot be synthesized accurately,
it also shows promising results that 
intrinsic MR contrast may be estimated from the remaining intrinsic contrasts.
%by the promising results in generating T1, T2, and T2F contrast in brain MR,
Thus,  we investigate whether
collaborative learning can essentially overcome the limitation of MAGiC images.
%Recall that MAGiC also calculates the voxel intensity from the multi-contrast MR images from the MDME scan. However, in contrast
%to MAGiC, the collaborative learning with CollaGAN also utilizes the semantic information beyond the pixel wise relationship, so the more systematic studies about the MR contrast can be performed, and we could obtain
%%Utilization of the multiple contrast images as inputs collaboratively, is a necessary step for the 
%accurate reconstruction of the desired MR contrast  
As shown in Fig.~\ref{fig:res_comparison} (a-b), accurate contrast was generated using CollaGAN by synergistically utilizing the redundancies across the remaining contrast.
In contrast to CycleGAN and StarGAN that utilize a single input MR image,  the accurate reconstructions of the voxel intensity are only possible by synergistically combining multiple contrast information via CollaGAN.
%Although CollaGAN produces excellent reconstruction results by finding the mapping from the multiple inputs collaboratively, the clinical evaluation of the results is more important since the motivation of CollaGAN was the clinical needs for synthesizing conventional T2-FLAIR images (T2-FLAIR*) using the other contrast MR images from MAGiC.
To verify the clinical efficacy of the method,
the reconstructed MR contrast images were undergone radiological evalution. % since it has systemic errors. 
The CollaGAN performs very well not only for the brain MR images from the normal subjects, but also for the brain scans from the subjects with lesions (Fig.~\ref{fig:res_lesion} (a-b)). 
The hyperintensity signal of the CSF space (yellow circled in Fig.~\ref{fig:res_lesion}(a)) compared to the other side hemisphere is well reconstructed on both MAGiC T2-FLAIR and T2-FLAIR.
Here, MAGiC T2-FLAIR and T2-FLAIR refer to the synthetic T2-FLAIR by MAGiC and the true T2-FLAIR contrast by additional acquisition, respectively.
The cortical and sulcal abnormality (yellow circled in Fig.~\ref{fig:res_lesion} (b)) are also visible on the reconstructed MAGiC T2-FLAIR and T2-FLAIR. The lesions of the subjects are well reconstructed compared to the original scans.
On the other hand, even if there exists systemic artifact on synthetic MAGiC T2-FLAIR, CollaGAN still reconstructs the artifact-free T2-FLAIR results with a help from the collaborative input as shown in Fig.~\ref{fig:res_lesion} (c-d).
The focal sulcal hyperintensity (yellow arrow in Fig.~\ref{fig:res_lesion} (c)) is only visible on T2-FLAIR (both original and reconstructed) while it is not visible on MAGiC  T2-FLAIR images.
Since the synthetic images (T1-FLAIR, T2-weighted, MAGiC T2-FLAIR) from MAGiC cannot capture the aforementioned hyper-intensity, it is usual to acquire additional scan of T2-FLAIR to detect the lesion. However, the hyper-intensity lesion could be detected on the reconstructed T2-FLAIR  by CollaGAN. % using only the synthetic images. 
%The synthetic T2-FLAIR cannot capture the hyperintensity lesion but CollaGAN can.
Also, in the reconstructed MAGiC T2-FLAIR, there exists a pseudo-lesion (yellow arrow in Fig.~\ref{fig:res_lesion}(d)) which is not visible on both original and reconstructed T2-FLAIR. 
%Thus, without CollaGAN, there should be additional T2-FLAIR scan to conclude that these candidates are real or not.
The radiologist concludes that the reconstructed conventional T2-FLAIR contrast  from CollaGAN not only reflects the original contrast well, but also removes the systemic artifacts from MAGiC well. 
By reconstructing the specific desired contrast MR images without any artifacts, we could save the scan time by avoiding the additional scan for accurate clinical diagnosis.
%\ref{fig:res_lesion})

%Therefore, we believe that CollaGAN is a promising algorithm for missing image data imputation in many real world applications.
In conclusion, we employ a  novel architecture, CollaGAN, to investigate the essential
MR contrast for imaging study, since CollaGAN can impute
missing image by synergistically learning the joint image manifold of multiple MR contrasts. 
%CollaGAN is free to scalability problem with the help of a single generator and a single discriminator. Also, CollaGAN handles the multiple inputs collaborately to generate the realistic and accurate reconstruction images.
%The multiple cycle consistency was introduced with the combinations of the inputs for stable training.
%%We showed that the proposed method produces images of higher visual quality compared to the existing methods.
%We showed that the proposed method produces images of high visual quality. 
Using the segmentation study, we found that images from the contrast agent are indispensible and cannot be completely
replaced by generative models.
For the case of intrinsic contrasts such as T2-FLAIR,
we demonstrated that CollaGAN reconstructs the specific contrast MR images without any artifacts, so that 
it can the scan time by avoiding the additional scan for accurate clinical diagnosis.
Our proposed CollaGAN model can be utilized for all other types of imaging studies to investigate which contrast is the most essential and
which contrast is redundant.

\section*{METHODS}
\subsection{Background Theory for CollaGAN}
%%%%%%
Here, we explain our Collaborative GAN framework to handle multiple inputs for generating more realistic  output for image imputation.
For ease of explanation, we assume that there are four types ($N=4$) of domains: $a$, $b$, $c$, and $d$.
To handle the multiple-inputs using a single generator, we train the generator to synthesize the output image in the target domain, $\hat{x}_a$, using a collaborative mapping from the set of the other types of multiple images, $\{x_a\}^C=\{x_b, x_c, x_d\}$,
where the superscript $^C$ denotes the complementary set.
This mapping is formally described by
\begin{eqnarray}
\hat{x}_\kappa= G\left(\{x_\kappa\}^C;\kappa\right) 
\end{eqnarray}
where $\kappa \in \{a,b,c,d\}$ denotes the target domain index that
 guides to generate the output of the proper target domain, $\kappa$. 
 As there are $N$ number of combinations for single-output and its corresponding complementary set as multiple-inputs,
  we randomly choose these combination during the training  so that the generator learns the various mappings to the multiple target domains. \\

One of the key concepts for the proposed method is the  {\it{multiple cycle consistency}}. 
Since the original cycle-consistency loss cannot be defined for the multiple inputs, the cyclic loss should be redefined.
Suppose that the {\it{fake}} output from the forward cycle for the generator, $G$, is $\hat{x}_a$.
%the auxilary set $X_{\hat{a}}=\{\hat{x}_a, x_b, x_c, x_d\}$ is defined. 
Then, we could generate $N-1$ number of new inputs by the combinations with the {\it{fake}} output, $\hat{x}_a$, and the inputs, $x_b, x_c, x_d$. Using the new combination inputs, the generator synthesizs the {\it{reconstructed}} outputs, $\tilde{x}_{\cdot|a}$, for the backward flow of the cycle. 
For example, when $N=4$,  there are three combinations of multi-input and single-output so that
we can reconstruct the three images of original domains using backward flow of the generator as:
\begin{eqnarray*}
\tilde{x}_{b|a} &=&G(\{\hat{x}_a, x_c,x_d\}; b) \\
\tilde{x}_{c|a} &=&G(\{\hat{x}_a, x_b,x_d\}; c) \\
\tilde{x}_{d|a} &=&G(\{\hat{x}_a, x_b,x_c\}; d) 
 \end{eqnarray*}
% and same for $\tilde{x}_c, \tilde{x}_d$.
% where $\kappa'=\{b,c,d\}$.
% and it is similar for $\tilde{x}_c$, and $\tilde{x}_d$. 
 Then, the associated multiple cycle consistency loss can be defined as following:
 \begin{eqnarray*}
  \mathcal{L}_{mcc, a} = ||x_b-\tilde{x}_{b|a}||_1 + ||x_c-\tilde{x}_{c|a}||_1 + ||x_d-\tilde{x}_{d|a}||_1 \ 
  \end{eqnarray*}
where $||\cdot||_1$ is the $l_1$-norm.
In general, the multiple cycle consistency loss for the multiple domains $\kappa$ can be written by
\begin{eqnarray}
 \mathcal{L}_{mcc,\kappa} = \sum_{\kappa'\neq \kappa} ||x_{\kappa'}-\tilde{x}_{\kappa'|\kappa}||_1
 \end{eqnarray}
where 
\begin{eqnarray}\label{eq:generate}
\tilde x_{\kappa'|\kappa} = G\left(\{\hat{x}_{\kappa}\}^C; \kappa'\right) \ .
\end{eqnarray}

To use a single generator, we need to use the mask vector to guide the generator to the target domain. The mask vector is an one-hot encoding vector which represents the target domain. When it is fed into the encoder part of $G$ (Fig.~\ref{fig:network}~ left), it is enlarged as same dimension with the input images to be easily concatenated. 
The mask vector has $N$ class number of channel dimension to represent the target domain as one-hot encoding along the channel dimension. This is a simplified version of mask vector which was originally introduced in StarGAN~\cite{choi2017stargan}.

\noindent{\bf Discriminator Loss}  As mentioned before, the discriminator has two roles: one is to classify the source which is real or fake, and the other is to classify the type of domain which is class $a, b, c$ or $d$. Therefore, the discriminator loss consists of two parts: adversarial loss and domain classification loss.
This can be realized using the two sub-paths $D_{gan}$ and $D_{clsf}$ in a single discriminator that shares the same neural network weights for feature extraction except the last layers for sub-paths.

Specifically, the adversarial loss is necessary to make the generated images as realistic as possible.
%the following least squares GAN loss\cite{mao2017least:
The regular GAN loss may lead to the vanishing gradients problem during the learning process~\cite{mao2017least,arjovsky2017wasserstein}. To overcome such problem and improve the robustness of the training, the adversarial loss of Least Square GAN~\cite{mao2017least} was utilized instead of the original GAN loss. 
In particular for the optimization of the discriminator, $D_{gan}$, the following loss is minimized:
$$\mathcal{L}_{gan}^{dsc}(D_{gan}) = \mathbb{E}_{x_\kappa} [(D_{gan}(x_\kappa)-1)^2]+\mathbb{E}_{\tilde x_{\kappa|\kappa}}[ (D_{gan}(\tilde{x}_{\kappa|\kappa}))^2],$$
whereas the generator is optimized by minimizing the following loss:
$$\mathcal{L}_{gan}^{gen}(G) = \mathbb{E}_{\tilde x_{\kappa|\kappa}}[ (D_{gan}(\tilde x_{\kappa|\kappa})-1)^2]$$
where $\tilde x_{\kappa|\kappa}$ is defined in \eqref{eq:generate}.

%\noindent{\bf Domain Classification Loss}
Next, the domain classification loss consists of two parts:  $\mathcal{L}_{clsf}^{real}$ and  $\mathcal{L}_{clsf}^{fake}$. They are the cross entropy loss for domain classification from the real images and the fake image, respectively.
Recall that the goal of training $G$ is to generate the image properly classified to the target domain. Thus, we first need a best classifier $D_{clsf}$ that should  be trained only with the real data to guide the generator properly. 
Accordingly, we first minimize the loss $\mathcal{L}_{clsf}^{real}$ to train the classifier $D_{clsf}$, then $\mathcal{L}_{clsf}^{fake}$ is minimized by training $G$ with fixing $D_{clsf}$ so that the generator can be trained to generate samples that can be classified correctly.

Specifically, to optimize the $D_{clsf}$, the following $\mathcal{L}_{clsf}^{real}$ should be minimizied with respect to $D_{clsf}$:
\begin{eqnarray}
\mathcal{L}_{clsf}^{real}(D_{clsf}) = \mathbb{E}_{x_{\kappa}} [-\log(D_{clsf}(\kappa;x_{\kappa}))]
%\sigma(o)^{(j)}]
\end{eqnarray}
where $D_{clsf}(\kappa;x_{\kappa})$ can be interpreted as the probability to correctly 
classify the real input $x_{\kappa}$ as the class $\kappa$.
%
%where $x_{\kappa'}=\{x_i|x_\kappa \not \in x_i\}$ is the $N$-1 number of images set X'except $x_{\kappa}$.
%%$$\mathbb{E}_{.} [-log(D_{clsf}(x_\kappa))]$$
%By minimizing $\mathcal{L}_{clsf}^{real}$, $D$ learns to classify a real image $x$ to its corresponding original domain $\kappa$. 
On the other hand, the generator $G$ should be trained to generate fake samples which are properly classified by the $D_{clsf}$. Thus, the following loss should be minimized with respect to $G$:
\begin{eqnarray}
\mathcal{L}_{clsf}^{fake}(G)=\mathbb{E}_{\hat x_{\kappa|\kappa}} [-\log(D_{clsf}(\kappa;\hat{x}_{\kappa|\kappa}))]
\end{eqnarray}

\noindent{\bf Structural Similarity Index Loss}  Structural Similarity (SSIM) index is one of the state-of-the-art metrics to measure the image quality~\cite{wang2004image}. 
The $l_2$ loss, which is widely used for the image restoration tasks, has been reported to cause the blurring artifacts on the results~\cite{ledig2017photo, mathieu2015deep, zhao2017loss}. SSIM is one of the perceptual metrics and it is also differentiable, so it can be backpropagated~\cite{zhao2017loss}. 
The SSIM for pixel $p$ is defined as
\begin{eqnarray}
\textrm{SSIM}(p) = \frac{2\mu_X\mu_Y+C_1}{\mu_X^2+\mu_Y^2+C_1} \cdot \frac{2\sigma_{XY}+C_2}{\sigma_X^2+\sigma_Y^2+C_2}
\label{eq:ssim}
\end{eqnarray}
where %$C_1, C_2, \mu_x, \mu_y, \sigma_x, \sigma_y, \sigma_xy$
$\mu_{X}$ is an average of $X$, $\sigma_{X}^2$ is a variance of $X$ and $\sigma_{X X^*}$ is a covariance of $X$ and $X^*$. 
There are two variables to stabilize the division such as $C_1=(k_1L)^2$ and $C_2=(k_2L)^2$.
$L$ is a dynamic range of the pixel intensities. $k_1$ and $k_2$ are constants by default $k_1=0.01$ and $k_2=0.03$. 
%
%%the dependence of menas and standard deviations on pixel $p$ was omitted.
%
%The Eq.~\ref{eq:ssim} refers that the computation of $SSIM(p)$ requires looking at a neighborhood of pixel $p$ as large as the support of Gaussian kernel. 
% Means and standard deviations are computed with a Gaussian filter with standard deviation $\sigma_G$, $G_{\sigma_{G}}$. 
 Since the SSIM is defined between 0 and 1, the loss function for SSIM can be written by:
\begin{eqnarray}
\mathcal{L}_{\textrm{SSIM}}(X,Y)=-\log\left( \frac{1}{2|P|}\sum_{p\in P(X,Y)}(1+\textrm{SSIM}(p))\right)
\end{eqnarray}
where $P$ denotes the pixel location set and $|P|$ is its cardinality.
The SSIM loss was applied as an additional multiple cycle consistency loss as follows:
\begin{eqnarray}
 \mathcal{L}_{mcc-\textrm{SSIM},\kappa} = \sum_{\kappa'\neq \kappa} \mathcal{L}_{\textrm{SSIM}}\left(x_{\kappa'},\tilde{x}_{\kappa'|\kappa}\right).
 \end{eqnarray}

%\noindent{\bf Least Sqaure Generative Adversarial Network} 

\subsection{Generator}
\begin{figure}%[h!]%  figure placement: here, top, bottom, or page
   \centering
\centerline{\epsfig{figure=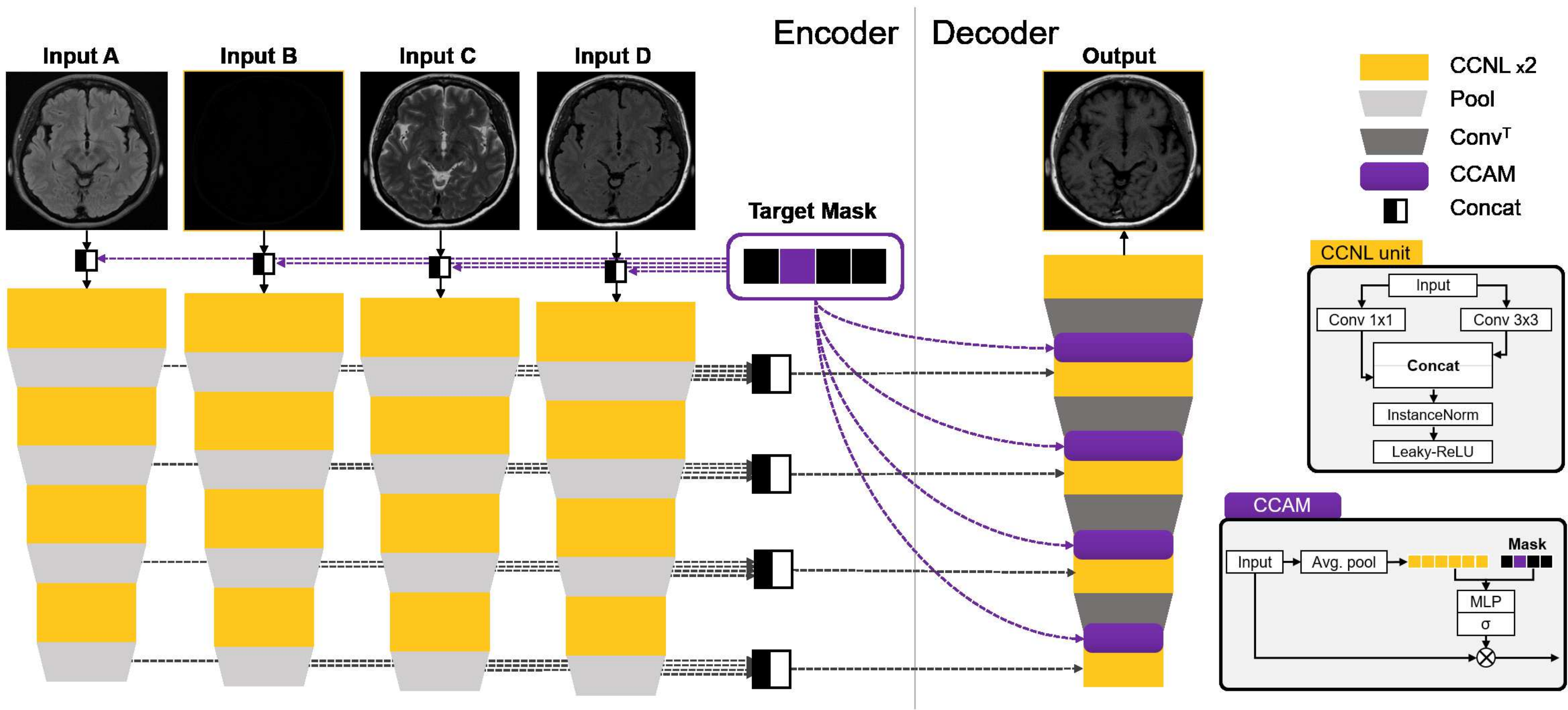, width=1\linewidth}}
   \caption{\bf\footnotesize 
   Architecture of the generator used for MR contrast imputation. 
The input images were concatenated with the target mask which represents the target domain. 
The network consists of CCNL unit (Concatenation of two Convolutions followed by instance Normalization and Leaky-relu, yellow), pooling layer (grey), convolution transpose layer (dark grey), concatenate layer (half balck/white squre), and CCAM (Conditioned Cahnnel Attention Module, purple). 
The network is divided by encoder part and decoder part. The encoder is multi-branched for individual feature extraction of each input MR contrast. 
The decoder part consists of series of CCNL units and CCAM. The CCAM gives the attention for accurate reconstruction of the contrast image in the target domain.   }
%The decoder structure shares with the discriminator used for the illumination translation. ($h\times w$)$\times N_{ch}$ represents the dimension of the features/images where $h$, $w$ and $N_{ch}$ is height, width and number of channels. The dashed arrow means skip connections. The downward, upward and right arrows represent [CNL$\times$2-P] layers, [T-CNL$\times$2] layers and [CNL$\times$2] layers, respectively, as explained in Table.~\ref{table:illum_G}.}
 \label{fig:network}
 \vspace*{0.5cm}
\end{figure}

CollaGAN consists of single pair of a generator, $G$, and a discriminator, $D$. For the generator, we redesigned U-net~\cite{ronneberger2015u} structure with the following three modification: CCNL (series of convolutions, cancatenation, normalization, and Leaky-ReLU layer) unit, multi-branched encoder, and channel attention as shown in Fig.~\ref{fig:network}.

First, the modified U-net basically consists of CCNL unit instead of CBR unit (series of convolution, batch normalization and ReLU layer) in original U-net architecture. Similar to the multi-resolution approach of GoogLeNet~\cite{szegedy2015going}, CCNL unit has two branched inputs: 1$\times$1 convolution and 3$\times$3 convolution layer. The two convolution layers are concatenated and pass through the Leaky-ReLU layer as shown in Fig.~\ref{fig:network}. It is important to utilize the 1$\times$1 convolution since the voxel-wise synthesis of the reconstruction is necessary as well as the 3$\times$3 convolution feature extraction for large receptive field. Thus, two branches of feature information are parallelly processed in CCNL units.

Second, we designed a multi-branched encoder for inidiviudal feature extraction for each input images (Fig.~\ref{fig:network} left). 
The generator consists of two parts: encoder and decoder. In the encoding step, each image are encoded separately by four branches. Here, the mask vector is concatenated to every input images to extract the proper features for the target domain. Then, the encoded features are concatenated at the end of the encoder and the concatenated features are fed into the decoder with the contracting paths between encoder and decoder. Since the inputs are not simply mixed in the first layer, the seperated features for each contrast image are extracted with a help of the multi-branched encoder. 

Third, the channel attention module called CCAM (Conditional Channel Attention Module)~\cite{chang2018image} is applied to the decoder part of the generator with the following modifications.
CCAM was originally designed for image translation to mixed-domain using Sym-parameterized Generative Network (SGN)~\cite{chang2018image}. 
CCAM selectively excludes channels and reduces influences of unnecessary channels to generate images in a mixed-domain conditioned by sym-parameters.
Here, we applied channel attention in the decoder part of the generator by CCAM modules using the one-dimensional mask vector as a sym-parameter. 
The input mask and the average pooled input features are concatenated and pass through the attention-MLP (multi-layered perceptron). The channel attentions are calculated as a form of scaling weights for each channel of input feature: 
$$CCAM(X,m) = X \cdot \sigma (MLP( [P_{avg}(X),m])) $$
where $X$ and $m$ represent the input features and 1-D input mask vector for target domain, respectively. And $P_{avg}$, $\sigma$ and $\cdot$ are the average pooling, the sigmoid operation and the elementwise multiplication, respectively.
The refined features are calculated by the element-wise multiplication between input features, $X$, and the scaling weights. 
The CCAM module chooses the channels with the calculated attention according to the target domain and the input features.

\subsection{Discriminator}

To classify the contrast of the MR images, the feature extraction by the multi-resolution processing is important. 
This kind of multi-scale approach is reported to work well in the  classification of MR contrasts~\cite{remedios10575classifying}.  
The discriminator has three branches that have different scales of resolution. 
Specifically, the first branch extracts the feature at the original scale of resolution and then reduces the size of feature domain. Another branch processes the feature extraction on the quater resolution scales ($height/4$, $width/4$). 
The other branch sequentially reduces the scales by two for extracting features. 
These three branches are concatenated to gather the features in multi-scale manner.
After that, the discriminator consists of three series of convolution with stride two and Leaky-ReLU.
At the end of the discriminator, there are two output headers: one is the source classification header for real or fake and the other is the domain classification header. PatchGAN~\cite{isola2017image, zhu2017unpaired} was utilized on the source classification header to classify whether local image patches are real or fake. We also found that the dropout~\cite{hinton2012improving, srivastava2014dropout} was very effective to prevent the overfitting of the discriminator.

\subsection{Brain Tumor segmentation datasets}
For quantitative analysis for the reconstruction performance, the multimodal brain tumor image segmentation benchmark (BraTS, 2015)~\cite{menze2015multimodal,bakas2017advancing} was used. BraTS supplies the routine clinically-acquired 3T multimodal MRI scans and the ground truth labels for brain tumor segmentation. The ground truth labels have been manually-revised by the expert board-certified neuroradiologists. 
The routine MRI scans consist of four different contrasts including native (T1), post-contrast T1-weighted (T1Gd), T2-weighted (T2), and T2-FLAIR (T2F) volumes, and were acquired with different clinical protocols and various scanners from multiple institutions. 
The datasets were divided into 218 / 28 / 28 subjects for train / validation / test sets, respectively.

\subsection{Tumor segmentation algorithm}
A semantic segmentation network for brain tumor segmentation from 3D MRIs using autoencoder regularization~\cite{myronenko20183d} achieved the top performance score in BraTS 2018 challenge. We implemented the segmentation network with some modifications to handle memory efficiently.

The segmentation network consists of shared encoder part and two branches of decoder part.
The encoder has an asymmetrically larger CNN architecture compared to the decoder part, to extract the features from the inputs. 
To fit into GPU memory size, we modified the 3D convolution layer to 2D convolution layer to perform 2.5D segmentation instead of 3D, which utilize the multiple neighborhood slices of MR images to map the single segmentation label. Here, we choose five slices (two adjacent slices from each dorsal and ventral slice) as input to find the tumor segmentation maps of the center slice.
The encoder part uses the blocks where each block consists of two convolutions with Group normalization (GN)~\cite{wu2018group} and ReLU, followed by additive identity skip connection. After the two unit blocks in each spatial level, the image dimensions were progressively downsized by two using the strided convolutions, and simultaneously increased feature size by two.

One branch of the decoder is for the segmentation map.
The decoder reconstructs each of the segmentation maps for following three tumor subregions: whole tumor, tumor core, and enhancing tumor core. The decoder utilised the same blocks in the encoder, but with a single block per each spatial level. The other branch of the decoder is for the regularization.  
The additional variational auto-encoder (VAE) branch reconstructs the input image itself to regularize the shared encoder during the training phase. 
The VAE branch was added to the encoder endpoint which is similar to auto-encoder architecture to additional guidance and regularization to the encoder part.

%Our comparison metric for the segmentation performance is dice similarity~\cite{dice1945measures}:$$ Dice(A,B) = \frac{2|A\cap B|}{|A|+|B|},$$where $|\cdot|$ represents the cardinalities of the set (number of elements in each set).

%Remarkable segmentation performance comes from the 

%We modified the architecture of the state-of-the-art convolutional neural network (CNN) for brain tumor segmentation~\cite{myronenko20183d}. 
%We modified the architecture of the state-of-the-art convolutional neural network (CNN) for brain tumor segmentation~\cite{myronenko20183d} by 2-D convolution layers instead of 3-D convolution layers because of limitation for GPU memory issue.
%

\subsection{Synthetic MR datasets}
We prepared the four types of contrasts for 280 axial brain images from 10 subjects. The subjects were scanned by the multi-dynamic multi-echo (MDME) sequence and the additional T2-FLAIR (FLuid-Attenuated Inversion Recovery) sequences.
Synthetic T1-FLAIR (T1F), T2-weighted (T2w) and MAGiC T2-FLAIR (T2F) images were acquired from MAGnetic resonance image Compilation (MAGiC, GE Healthcare)~\cite{tanenbaum2017synthetic} using the MDME scans. 
The MR scan parameters for T1-FLAIR / T2-weighted / MAGiC T2-FLAIR are as followings: TR 2500ms, TE 10ms, TI 1050ms, FA 90$\degree$ / TR 3500ms, TE 128ms, FA 90$\degree$ / TR 9000ms, TE 95ms, TI 2408ms, FA 90$\degree$, respectively.
The additional T2-FLAIR  scans was acquired with different scan parameter of T2F: TR 9000ms, TE 93ms, TI 2471ms, FA 160$\degree$. The followings are common parameters for four scans: FOV 220$\times$220mm, 320$\times$224 acquisition matrix, 4.0mm slice thickness.
The MR images were divided into the train (224 images), validation (28 images) and test sets (28 images) by the subjects.

\subsection{Data preprocessing and Augmentation}
The MR images were normalized to have unit standard deviations based on the non-zero-voxels only. For the data augmentation, we apply a random scale (0.9-1.1) and a random flip on lateral-to-lateral direction with a probability 0.5.

%Soft Dice Loss~\cite{milletari2016v}

\clearpage

%\newpage

%\bibliography{egbib}

\bibliographystyle{naturemag}

 \begin{addendum}
 \item This research was  supported by the National Research Foundation (NRF) of Korea  grant  NRF-2016R1A2B3008104.

%\item[Author Contributions] J.C.Y. supervised the project in conception and discussion. J.M., K. K, J.J., and J.C.Y. designed the experiments and analysis. J. M, K.K, and  J.M. performed all experiments and analysis and prepared nanoparticles sample. S.-W.R and C.C. prepared and analyzed all biological samples. D.K. and K.-H.J. designed and prepared nanopatterns sample.  J.M., K. K, J.J., and J.C.Y wrote the manuscript.
 \item[Competing Interests] The authors declare that they have no competing financial interests.
 %\item[Correspondence] Correspondence and requests for materials should be addressed to Jong Chul Ye.~(email: jong.ye@kaist.ac.kr).
\end{addendum}

\clearpage

%\begin{figure}[h]
%	\centering
%	\includegraphics[width=12cm]{./fig1-1.jpg}
%	%\captionsetup{justification=raggedright,singlelinecheck=false}
%	\caption{
%		Reconstruction flow for (a) ALOHA-based ghost artifact removal, and (b) the proposed $k$-space deep learning for ghost artifact removal. Here, $e$ and $o$ refer to the frames composed of the even and odd $k$-space lines, respectively. After the interpolated even and odd virtual images are generated, the sum-of-squares image is obtained as the final ghost corrected image. Here, IFT stands for the inverse Fourier transform. }
%	\label{fig:overview}
%\end{figure}

%\begin{figure}[!h]
%\centerline{\epsfig{figure=figs/eps/flow_colla.eps, width=1\linewidth}}
%   \centering
%   \caption{\bf\footnotesize 
%Flow of CollaGAN. The generator, $G$, reconstructs the target domain MR contrast using the set of input MR contrasts (left). For the multiple cycle consistency, the generated fake image re-entered to $G$ with inputs images and $G$ produces the multiple reconstructed outputs correspond to the original domains.
%$D$ has two branches : domain classifier, $D_{clsf}$  and source classfier, $D_{gan}$ (real/fake) . $D_{clsf}$ and $D_{gan}$ shares the layers for feature extractions.
%Here, $D_{gan}$ and $D_{clsf}$ are simultaneously trained by the loss from both (1) fake \& (2) real images and only (2) real images, respectively.    
%}
%
%   \vspace*{0.5cm}
%\end{figure}
%   \label{fig:flow_colla}

%Image translation tasks for multi-domain (a-c). (a) Cross-domain model needs large number of generators and discriminaors to handle multi-domain data. (b) StarGAN uses a single generator with one input from single domain. (c) 

%The original T2-FLAIR* has structural difference compared to the original T1-FLAIR, T2-weighted, and T2-FLAIR since the original T2-FLAIR* is the separated scan from the others and the movement of the patient might exist.

\printfigures

\end{document}